\documentclass[]{amsart}
\usepackage{amsmath}
\usepackage{biblatex}
\usepackage{algorithm}
\usepackage{algpseudocode}
\usepackage{algorithmicx}
\usepackage{booktabs}

\begin{document}
\title{A path-finding algorithm for computing minimal-weight-matching centrosymmetry parameter}

\author[]{Vasily V. Pisarev}
\address{HSE University Tikhonov MIEM, Moscow, Russia}
\email{v.pisarev@hse.ru}

\maketitle

\begin{abstract}
In 2020, Peter Larsen has noticed flaws in the methods for centrosymmetry parameter
computation in the existing molecular dynamics and analysis packages. He proposed
an intuitive an mathematically rigorous formulation for centrosymmetry parameter
in terms of minimal-weight matching (MWM) on a fully-connected graph of atomic neighbors.
He proposed using Edmonds' blossom algorithm for computing such a matching.
In this paper, we investigate an alternative algorithm for MWM CSP computation
using path finding approach and A* algorithm.
\end{abstract}

\section{Centrosymmetry parameter}

The centrosymmetry parameter has been defined as
\begin{equation}
\text{CSP} = \sum\limits_{i=1}^{N/2} ||\mathbf{R}_i + \mathbf{R}_{i+N/2}||^2,
\end{equation}
\noindent where $N$ is the number of nearest neighbors considered for CSP computation (an even number), $\mathbf{R}_i$ and $\mathbf{R}_{i+N/2}$ are the vectors corresponding to the pairs of opposite neighbors in the lattice.

The choice of specific pairs has not been described. Larsen~\cite{Larsen2020_arxiv} reports of two variants of the choice, greedy edge selection and greedy vertex matching, prevailing in the software by 2020, each of which has serious drawbacks. The reader is invited to read Larsen's paper for further details on those two methods.

Larsen proposed reformulation of the edge selection problem for CSP as a constrained discrete optimization problem:
\begin{enumerate}
\item Each atom must participate in exactly one edge, and
\item The sum of edge weights must be minimal.
\end{enumerate}

Those two conditions constitute a minimal-weight maximum matching problem on a fully-connected graph of atoms where weight $w_{ij}$ of the edge connecting atoms $i$ and $j$ equals to $||\mathbf{R}_i + \mathbf{R}_j||^2$.

\section{An alternative formulation}

Let us consider a graph where nodes are all possible matchings on the atom graph, edges connect each $k$-pair
matching $M_{p_1, \dots, p_k}$ with all $k+1$-pair matchings $M_{p_1, \dots, p_k, p_{k+1}}$ formed by
adding another pair of atoms to $M_{p_1, \dots, p_k}$ and the weight of the edge equals to the weight of the added atom pair. Then, finding the minimal-weight maximum matching is finding the shortest path from a 0-pair matching to a $N/2$-pair matching in that graph. As a concrete example, A* algorithm finds a minimum path and has optimal time complexity of all heuristic path search algorithms Although the time complexity for the blossom algorithm is $O(N^4)$ and the worst-case runtime of A* algorithm is $O(b^{N/2})$ where $b$ is the branching factor which can be conservatively estimated as $O(N^2)$, there are a few reasons to consider a path search reformulation for the specific problem:
\begin{itemize}
\item CSP calculations are done for a small number of neighbors, typically 8 for BCC first shell, 12 for FCC first shell or 14 for BCC second shell
\item The A* algorithm is more straightforward than blossom algorithm, so that the smaller constant factors may outweigh the larger big-O complexity
\item With a suitable heuristic and early pruning, effective branching factor can be much smaller than the worst-case estimate.

\end{itemize}

\section{Algorithm}

The general A* search algorithm is defined as follows~\cite{Hart1968}.

\begin{algorithm}
\caption{A* Path Search}
\begin{algorithmic}[1]
\Function{Astar}{$G, W, v_\text{start}, v_\text{goal}, h$}
\State \textbf{input:} graph $G=\{\text{vertices} \{v_k\}, \text{edges} \{e_l\}\}$, edge weight matrix $W$, starting vertex $v_\text{start}$, goal vertex $v_\text{goal}$, heuristic $h(v)$
\State \textbf{output:} A path $S=\{e_\text{path}\}$ from $v_\text{start}$ to $v_\text{goal}$
\State $Q \gets \Call{min\_heap}{\text{}}$ \Comment{Min-heap ordered by tentative f-score}
\State $\Call{enqueue}{Q, [v_\text{start}]}$
\While{$Q \neq \emptyset$}
    \State $s \gets \Call{dequeue}{Q}$
    \State $v_\text{current} \gets \Call{last\_vertex}{s}$
    \If{$v_\text{current} = v_\text{goal}$}
        \State \Return $s$ \Comment{Goal reached}
    \EndIf
    \For{$v_\text{next} \in \Call{neighbors}{G, v_\text{current}}$}
        \State $s' \gets \Call{append}{s, v_\text{next}}$
        \State $s'.\texttt{f\_score} \gets \Call{path\_cost}{s} + W[v_\text{current}, v_\text{next}] + h(v_\text{next})$
        \State $\Call{enqueue}{Q, s'}$
    \EndFor
\EndWhile
\State \Call{ERROR}{$\text{}$} \Comment{No more vertices to explore but path has not been found}
\EndFunction
\end{algorithmic}
\end{algorithm}

If the heuristic $h(v)$ is admissible, i.e. does not overestimate true path cost from $n$ to $v_\text{goal}$, then the A* algorithm finds the shortest path, i.e. it provides a valid solution for our specific problem. However, Algorithm 1 as written is not competitive for the CSP computation due to large work needed for new path exploration. In the rest of the paper, we explore possible optimizations tailor-fitted for the specific problem.

\subsection{Path representation}

Each node in our graph represents a \emph{partial matching} -- a set of disjoint atom pairs. If each of the atom pairs has a unique numeric identifier (ID), then each partial matching has a unique representation in which all pairs are sorted in the ascending order by the ID. Therefore, a path from an emty matching to the matching $M_{p_1, \dots, p_k}$ is represented by the list $\{p_1, \dots, p_k\}$ sorted by ID. Then, to form the next matching, we can consider only adding pairs with ID from $\text{ID}_k + 1$ onwards. Furthermore, for CSP calculation we are not interested in the set of pairs itself, only in its total weight. In that case, the state representation is further reduced. In the proposed approach, a state is characterized by:
\begin{itemize}
    \item The set of atoms already matched, represented as a bitmask $m$. The number of atoms for CSP calculation does not exceed 16, so that standard 16-, 32- or 64-bit integers are suitable for that purpose,
    \item The largest pair identifier in the matching, $\text{ID}_\text{max}$,
    \item The accumulated weight of the matching so far,
    \item The expected total weight, which includes a lower bound on the cost to complete the matching.
\end{itemize}

The search begins from the empty matching (0-pair state) and terminates upon reaching a complete matching ($N/2$-pair state where all atoms are covered). Edges in the state space graph correspond to adding one atom pair to the current partial matching. The weight of a transition equals the weight of the added pair, $w_{ij} = ||\mathbf{R}_i + \mathbf{R}_j||^2$. If the pairs are represented as bitmasks of connected atoms, then the next possible pair to add can be found efficiently by linear scan over the list of pairs and the check of bitwise $m \text{ }\texttt{\&} \text{ } ij$.

\subsection{Path queue representation}

The set $Q$ of next nodes to consider in A* is implemented as a priority queue ordered by expected path weight $f = g + h$, implemented as a standard binary heap.

\subsection{Pair ordering}

For the efficient implementation, the weights of all atom pairs are computed in advance and sorted in the ascending order. Firstly, it allows us to use the greedy edge selection estimate. Secondly, it allows further optimizations as described below. The weights are stored in an array which has a length $N_\text{pairs} = N(N-1)/2$.

\subsection{Path length heuristic}
\label{sec:heuristic}

The A* algorithm requires a heuristic function $h(M)$ estimating the cost to complete a partial matching $M$. We define this heuristic as the sum of the $r$ shortest edges with IDs larger than $\text{ID}_\text{max}$ of the matching, where $r = N/2 - |M|$ is the number of additional pairs needed.

The advantage of this heuristic is that it only depends on the last matched pair, and the values can be statically precomputed prior to A* execution. It is computed in the form of $N_\text{pairs} \times N/2$ matrix $H$ where row number represents $\text{ID}_\text{max}$ and column number represents the remaining number of pairs $r$.
\begin{equation}
    H(p, r) = \sum_{k=p+1}^{p+r} w_{(k)},
\end{equation}
where $w_{(k)}$ denotes the weight of the $k$-th pair in the sorted list. If $p+r > N_\text{pairs}$, then the corresponding element of the matrix is set to infinity or to a large enough value.

For a state with last-added edge at index $\text{ID}_{\text{max}}$ and $r$ remaining pairs to add, the heuristic value is $h = W_{\min}(k_{\text{max}}, r)$. The expected total weight is then $f = g + h$, where $g = w_{\text{path}}$ is the actual cost so far.

\subsubsection{Properties of the heuristic}

\textbf{Admissibility}

A heuristic is admissible if it never overestimates the true cost to reach the goal. Our heuristic is admissible because:

\begin{enumerate}
    \item The sum $H(p, r)$ considers the $r$ minimum-weight atom pairs starting from index $p+1$, ignoring any conflicts between them.
    \item Any valid completion of the partial matching must select $r$ pairs, starting from index $p+1$, which do not overlap with the pairs already present in the matching
    \item Therefore, all pairs on the true path completing a matching $M$ have weights larger or equal than the pairs used to compute $H(p, r)$, and $H(p, r) \leq h^*(M)$, where $h^*(M)$ is the true minimum cost to complete matching $M$.
\end{enumerate}

The admissibility is guaranteed by sorting atom pairs in the ascending-weight order prior to A* execution.

\textbf{Monotonicity}

Each column of $H$ is sorted in the ascending order:
\begin{equation}
 H(p, r) = \sum_{k=p+1}^{p+r} w_{(k)} = H(p-1, r) - w_{(p)} + w_{(p+r)} \geq H(p-1, r).
 \label{eq:h_monotonic}
\end{equation}

Therefore, if $H(p, r) > C$, then $H(p+i, r) > C$ for all $i \geq 1$.

\subsection{A* algorithm optimized for CSP}

Assuming that the set of covered atoms $M$ is stored as 32-bit integer and \texttt{popcnt} is a function returning the number of nonzero bits in the binary number reprezentation, the algorithm can be written as follows.

\begin{algorithm}
\caption{A* algorithm optimized for CSP: No early pruning}
\label{alg:astar_csp}
\begin{algorithmic}[1]
\Function{MWM\_CSP\_Astar}{$R$}
\State \textbf{input:} array of particle positions $R$
\State \textbf{output:} minimum-weight matching CSP $C_m$
\State $N \gets \Call{length}{R}$
\State $N_\text{pairs} \gets N(N-1)/2$
\State $Pp \gets \Call{sort}{[(||{R}_i + {R}_j||^2, \texttt{1<<i | 1<<j}) \text{ for } i \in \{0,\dots, N-1\}, j \in \{i+1, \dots, N-1\} ]}$
\State $H \gets \Call{generate\_h\_matrix}{Pp, N}$
\State $w_\text{est} \gets Pp[1][1] + H[1, N/2-1]$ \Comment{greedy path weight estimate}
\State $Q \gets \Call{min\_heap}{\text{}}$ \Comment{Min-heap ordered by tentative f-score}
\State $\Call{enqueue}{Q, (w_\text{est}, 0.0, 0, 0)}$
\While{$Q \neq \emptyset$}
    \State $w_\text{est}, g, M, \text{ID}_\text{max} \gets \Call{dequeue}{Q}$
    \State $r \gets (N - \Call{popcnt}{M}) / 2$
    \If{$r = 0$}
        \State \Return $w_\text{est}$ \Comment{Goal reached}
    \EndIf
    \For{$\text{ID} \in \text{ID}_\text{max}+1, \dots, N_\text{pairs}$}
        \State $(w, ij) \gets Pp[\text{ID}]$
        \If{$(M \texttt{ \& } ij) = 0$} \Comment{Next selected edge is not in $M$}
            \State $w_\text{est} \gets g + w + H[\text{ID}, r]$
            \State $\Call{enqueue}{Q, (w_\text{est}, g + w, M \texttt{ | } ij, \text{ID})}$
        \EndIf
    \EndFor
\EndWhile
\State \Call{ERROR}{$\text{}$} \Comment{No more vertices to explore but path has not been found}
\EndFunction
\end{algorithmic}
\end{algorithm}

Algorithm~\ref{alg:astar_csp} assumes 1-based indexing for arrays and tuples, \texttt{|} and \texttt{\&} are used for bitwise ``or'' and ``and'', respectively, \texttt{<<} is used for bitwise shift. generate\_h\_matrix is a function filling the heuristic matrix as explained in Sec.~\ref{sec:heuristic}.

\section{Early-pruning strategies}

Although Algorithm~\ref{alg:astar_csp} correctly finds the minimum-weight path, further optimizations are possible. Profiling shows that the majority of time is spend in priority queue operations, so that additional optimizations are aimed to optimize them by pruning the branches that are guaranteed to not yield the optimal solution before they are enqueued.

\begin{enumerate}
    \item \textbf{Initial upper bound}: A greedy vertex matching (selecting the shortest non-conflicting edges) provides an initial upper bound $w_U$ on the optimal MWM weight. Any path with $w_\text{est} > U$ cannot improve the solution.

    \item \textbf{Dynamic upper bound}: During search, when a complete matching is discovered, $U$ is updated to the minimum of its current value and the new matching's weight. This tightens pruning criteria as the search progresses.

    \item \textbf{Branch pruning}: Due to heuristic monotonicity property~\eqref{eq:h_monotonic}, once an pair $k$ satisfies $g + w_{(k)} + H[k, r] > w_U$, all subsequent pairs also satisfy this condition. Therefore, the inner loop in Algorithm~\ref{alg:astar_csp} can be terminated once it is met, because no other next pair can improve the MWM weight estimate.

    \item \textbf{Queue compaction}: As upper bound $w_U$ is updated, the queue is resized to the index of the last item for which $w_\text{est} \leq w_U$. That procedure further restricts the size of the queue.

    \item If the number of remaining pairs with $\text{ID} > \text{ID}_\text{max}$ is insufficient to complete the matching, the state is not expanded.
\end{enumerate}

These strategies are included in Algorithms~\ref{alg:astar_csp_pruning} and \ref{alg:astar_csp_init}.

\begin{algorithm}
\caption{A* algorithm optimized for CSP with pruning strategies}
\label{alg:astar_csp_pruning}
\begin{algorithmic}[1]
\Function{MWM\_CSP\_Astar}{$R$}
\State \textbf{input:} array of particle positions $R$
\State \textbf{output:} minimum-weight matching CSP $C_m$
\State $N \gets \Call{length}{R}$
\State $N_\text{pairs} \gets N(N-1)/2$
\State $Pp \gets \Call{sort}{[(||{R}_i + {R}_j||^2, \texttt{1<<i | 1<<j}) \text{ for } i \in \{0,\dots, N-1\}, j \in \{i+1, \dots, N-1\} ]}$
\State $(w_\text{greedy}, w_U) \gets \Call{initial\_weight\_guesses}{Pp}$
\If{$w_\text{greedy} = w_U$} \Comment{Greedy search gives the needed matching}
    \State \Return $w_\text{est}$
\EndIf
\State $H \gets \Call{generate\_h\_matrix}{Pp, N}$
\State $Q \gets \Call{min\_heap}{\text{}}$ \Comment{Min-heap ordered by tentative f-score}
\State $\Call{enqueue}{Q, (w_\text{greedy}, 0.0, 0, 0)}$
\While{$Q \neq \emptyset$}
    \State $w_\text{est}, g, M, \text{ID}_\text{max} \gets \Call{dequeue}{Q}$
    \State $r \gets (N - \Call{popcnt}{M}) / 2$
    \If{$r = 0$}
        \State \Return $w_\text{greedy}$ \Comment{Goal reached}
    \EndIf
    \State $w_U' \gets g$ \Comment{Initialize values for upper bound guess}
    \State $M' \gets M$
    \For{$\text{ID} \in \text{ID}_\text{max}+1, \dots, N_\text{pairs}$}
        \State $(w, ij) \gets Pp[\text{ID}]$
        \State $w_\text{est} \gets g + w + H[\text{ID}, r]$
        \If{$w_\text{est} < w_U$}
            \If{$(M \texttt{ \& } ij) = 0$} \Comment{Next selected edge is not in $M$}
                \State $\Call{enqueue}{Q, (w_\text{est}, g + w, M \texttt{ | } ij, \text{ID})}$
            \EndIf
            \If{$(M' \texttt{ \& } ij) = 0$}
                \State $M' \gets (M' \texttt{ | } ij)$
                \State $w_U' \gets w_U' + w$
            \EndIf
        \Else \State \textbf{break}
        \EndIf
    \EndFor
    \If{$\Call{popcnt}{M'} = N$ \textbf{and} $w_U' < w_U$} \Comment{Found a complete matching with better weight estimate}
        \State $w_U \gets w_U'$
        \State $\Call{compact\_queue}{Q, w_U}$
    \EndIf
\EndWhile
\State \Return $w_U$ \Comment{No more vertices to explore, upper estimate is the best}
\EndFunction
\end{algorithmic}
\end{algorithm}

\begin{algorithm}
\caption{Initial guesses for lower and upper bounds}
\label{alg:astar_csp_init}
\begin{algorithmic}[1]
\Function{initial\_weight\_guesses}{$Pp$}
\State \textbf{input:} sorted array of pairs $Pp$
\State \textbf{output:} lower bound $w_\text{greedy}$, upper bound $w_U$
\State $M \gets 0$
\State $w_U = 0.0$
\State $w_\text{greedy} \gets 0.0$ \Comment{greedy path weight estimate}
\State $n_\text{greedy} \gets 0$ \Comment{Upper-bound weight estimate}
\For{$(w, ij) \in Pp$}
    \If{$n_\text{greedy} < N/2$}
        \State $w_\text{greedy} \gets w_\text{greedy} + w$
        \State $n_\text{greedy} \gets n_\text{greedy} + 1$
    \EndIf
    \If{$(M \texttt{ \& } ij) = 0$}
        \State $w_U \gets w_U + w$
        \State $M \gets (M \texttt{ | } ij)$
        \If{$\Call{popcnt}{M} = N$}
            \State \textbf{break}
        \EndIf
    \EndIf
\EndFor
\State \Return $(w_\text{greedy}, w_U)$
\EndFunction
\end{algorithmic}
\end{algorithm}

The early return in Line 8 of MWM\_CSP\_Astar when greedy edge selection finds a true solution is not strictly necessary because the initial path would still cause an early termination of the A* loop in that case. However, checking it early still saves some runtime in easy cases by avoiding the $H$ matrix generation.

\section{Implementation}

Algorithm~\ref{alg:astar_csp} has been implemented in Julia language within the MDProcessing.jl package. It has been translated to C++ and benchmarked against the reference blossom algorithm implementation by P. Larsen and D. Pereira available in OVITO package~\cite{ovito_mwm_csp}.

\subsection{Benchmarks}

The C++ source code for the reference MWM CSP implementation and the proposed one have been isolated to be called from MDProcessing.jl package using the precomputed neighbor distance buffers. As such, the differences in runtime of two C++ implementations are only due to algorithmic differences as the Julia runtime overhead includes the same operations in both cases. The Julia implementation is also called from MDProcessing.jl, the only difference is that it uses preallocated buffers for atom pairs, path queue and $H$ matrix.

The single-thread runtimes are presented for all MWM CSP implementations on a desktop and a laptop computers under openSUSE operating system with Linux 7.0 kernel. The C++ implementations are called from shared libraries compiled with \texttt{-Ofast} flag using GCC 15.2 compiler. Julia implementation uses Julia 1.11 version of the runtime. The averaged runtimes and standard uncertainties are evaluated using BenchmarkTools.jl~\cite{benchmarktools}.

The first benchmark is a system of 1000 Lennard-Jones particles at reduced density $\rho^* = 0.7$ and reduced temperature $T^* = 1.0$, corresponding to the liquid phase. It corresponds to the case when greedy edge selection fails and full MWM computation is used for most of the particles.

\begin{table}[htbp]
\centering
\caption{Runtime comparison for different implementations and hardware: LJ liquid}
\label{tab:execution_time}
\begin{tabular}{l p{0.3\linewidth} p{0.3\linewidth}}
\toprule
Run parameters & Time (ms), Laptop (Ryzen 5 3500U, DDR4-2400) & Time (ms), Desktop (Ryzen 7 7700, DDR5-3400) \\
\midrule
N = 8, C++ blossom & $10.7 \pm 0.2$ & $4.7 \pm 0.1$ \\
N = 8, C++ A* & $3.8 \pm 0.3$ & $2.16 \pm 0.04$ \\
N = 8, Julia A* & $3.6 \pm 0.2$ & $2.14 \pm 0.05$ \\
N = 12, C++ blossom & $27.8 \pm 3.3$ & $10.8 \pm 0.1$ \\
N = 12, C++ A* & $10.8 \pm 0.8$ & $6.6 \pm 0.08$ \\
N = 12, Julia A* & $10.9 \pm 1.2$ & $6.6 \pm 0.07$ \\
N = 14, C++ blossom & $35.9 \pm 2.9$ & $15.5 \pm 0.1$ \\
N = 14, C++ A* & $22.4 \pm 1.1$ & $13.9 \pm 0.06$ \\
N = 14, Julia A* & $23.1 \pm 2.1$ & $14.1 \pm 0.1$ \\
N = 16, C++ blossom & $50.6 \pm 2.7$ & $21.6 \pm 0.1$ \\
N = 16, C++ A* & $53.0 \pm 0.9$ & $33.7 \pm 0.16$ \\
N = 16, Julia A* & $55.8 \pm 4.0$ & $34.6 \pm 0.15$ \\
\bottomrule
\end{tabular}
\end{table}

The table shows that, although A* algorithm scales worse than the blossom algorithm for computing the minimum-weight matching, with additional optimization the latter is faster for small graph sizes typical for CSP calculation. The break-even point is between 14 and 16 atoms. Given that the CSP computation is usually done for 8-14 atoms, the A* search is worth considering as the default MWM implementation.

The Julia implementation is competitive against the C++ version, being only a few percent slower.

The second benchmark analyzes runtime for a crystalline system of 8000 Cu atoms. In that case, 12-atom matching must proceed faster due to high rate of greedy edge selection success. Table~\ref{tab:execution_time_cu} shows the results. As expected, the original implementation with GES first and blossom algorithm as fallback performs the best with 12 atoms. For other numbers of neighbors, the fallback method is activated, and the runtimes are larger. For the neighbor counts different than 12, the A* algorithm is faster in all cases. For 12 neighbors, A* is 20\% slower on the desktop system, likely due to extra overhead of initial sorting of the edges array. A potential further optimization is to keep greedy edge selection as implemented by Larsen and use the algorithm described here only as fallback.

\begin{table}[htbp]
\centering
\caption{Runtime comparison for different implementations and hardware: Cu crystal}
\label{tab:execution_time_cu}
\begin{tabular}{l p{0.3\linewidth} p{0.3\linewidth}}
\toprule
Run parameters & Time (ms), Laptop (Ryzen 5 3500U, DDR4-2400) & Time (ms), Desktop (Ryzen 7 7700, DDR5-3400) \\
\midrule
N = 8, C++ blossom & $73.7 \pm 2.5$ & $31.6 \pm 0.2$ \\
N = 8, C++ A* & $24.3 \pm 2.4$ & $13.9 \pm 0.2$ \\
N = 8, Julia A* & $23.2 \pm 0.8$ & $13.4 \pm 0.08$ \\
N = 12, C++ blossom & $33.4 \pm 3.1$ & $16.0 \pm 0.1$ \\
N = 12, C++ A* & $33.4 \pm 1.6$ & $19.9 \pm 0.2$ \\
N = 12, Julia A* & $31.6 \pm 0.3$ & $19.4 \pm 0.1$ \\
N = 14, C++ blossom & $172.5 \pm 8.6$ & $80.1 \pm 0.5$ \\
N = 14, C++ A* & $68.7 \pm 4.1$ & $39.9 \pm 0.2$ \\
N = 14, Julia A* & $73.1 \pm 4.2$ & $39.6 \pm 0.3$ \\
N = 16, C++ blossom & $278.5 \pm 9.3$ & $115.6 \pm 0.5$ \\
N = 16, C++ A* & $100.3 \pm 0.8$ & $64.1 \pm 0.4$ \\
N = 16, Julia A* & $106.9 \pm 0.6$ & $65.2 \pm 0.3$ \\
\bottomrule
\end{tabular}
\end{table}

\section{Conclusions}

An alternative implementation of minimal-weight matching for centrosymmetry parameter is proposed. It is based on direct graph search using A* algorithm with extra pruning and early-termination conditions to avoid exploration of dead-ends.

The proposed algorithm is competitive with the standard blossom algorithm proposed earlier for this problem if the number of atoms is 16 or less. As such, the method is worth considering as the default for centrosymmetry calculations in atomistic simulations.

\printbibliography

\end{document}